\documentclass[aps,prl]{revtex4}
\usepackage{graphicx}
\begin{document}
\title{Elastic properties of proteins: insight on the folding process
and evolutionary selection of native structures}

\author{Cristian Micheletti}
\affiliation{International School for Advanced Studies (S.I.S.S.A.) and INFM, \\
\small Via Beirut 2-4, 34014 Trieste, Italy}
\author{Gianluca Lattanzi}
\affiliation{Hahn-Meitner Institut, Abt. SF5, \\
\small Glienicker Str. 100, 14109 Berlin, Germany}
\author{Amos Maritan}
\affiliation{International School for Advanced Studies (S.I.S.S.A.) and INFM, \\
\small Via Beirut 2-4, 34014 Trieste, Italy \\
The Abdus Salam International Center for Theoretical Physics, \\
\small Strada Costiera 11, 34100 Trieste, Italy}
\date{\today}
\vskip 0.5cm
\begin{abstract}
We carry out a theoretical study of the vibrational and relaxation
properties of naturally-occurring proteins with the purpose of
characterizing both the folding and equilibrium thermodynamics. By
means of a suitable model we provide a full characterization of the
spectrum and eigenmodes of vibration at various temperatures by merely
exploiting the knowledge of the protein native structure. It is shown
that the rate at which perturbations decay at the folding transition
correlates well with experimental folding rates. This validation is
carried out on a list of about 30 two-state folders. Furthermore, the
qualitative analysis of residues mean square displacements (shown to
accurately reproduce crystallographic data) provides a reliable and
statistically accurate method to identify crucial folding
sites/contacts. This novel strategy is validated against clinical data
for HIV-1 Protease. Finally, we compare the spectra and eigenmodes of
vibration of natural proteins against randomly-generated compact
structures and regular random graphs. The comparison reveals a
distinctive enhanced flexibility of natural structures accompanied by
slow relaxation times at the folding temperature. The fact that these
properties are intimately connected to the presence and assembly of
secondary motifs hints at the special criteria adopted by evolution in
the selection of viable folds.
\end{abstract}
\maketitle

\section{Introduction}

The backbone structure of globular proteins displays a notable degree
of order and organization resulting in secondary motifs such as
$\alpha-$helices and $\beta-$sheets and in their optimal arrangement
into a compact shape. The
special neatness and regularity of native conformations impacts on, or
rather reflects, several features that are distinctive of naturally
occurring proteins
\cite{bakernature,opthelix,Miche99b,secstr}. Indeed, experimental and
theoretical studies have shown that native structures conceal a wealth
of information about aspects as diverse as native elastic properties,
folding rates and key folding stages
\cite{bav93,Plaxco,Chan98,hiv}. 
A promising graph theoretical approach~\cite{jacobs01,rader02} can be used to 
ifentify rigid and flexible regions in proteins, and to investigate the 
emergence of such flexible regions in unfolding processes.
Recent topology-based approaches,
which are independent of sequence specificity, have revealed that a
schematic, coarse-grained structural description is sufficient to
obtain results in remarkable qualitative and even quantitative
agreement with known experimental facts.

For example, concerning near-native vibrations, one is entitled to
replace the detailed interactions among amino acids which are in
contact in the native state with springs acting on effective centroids
(usually the $C_\alpha$ atoms)~\cite{bav93,tir96}. Topological folding
models, on the other hand, try to characterize how the loss of
configurational entropy contrasts the progressive establishment of
native interactions in a folding process
\cite{Go,Miche99b,Finkel,eaton,baker,thirum,Clem,hiv}.

In the present study, we focus on a topology-based model that
incorporates both the aspects mentioned above. It consists of a
beads-and-springs model but, at variance with other approaches, the
strength of the effective interaction depends on the temperature of
the thermal bath~\cite{gaussian}. Since the model is amenable to
analytic treatment, it is possible to characterize rigorously both the
thermodynamics as well as the vibrational/relaxation dynamics at
various equilibrium stages of the folding process (not only near the
native state).

The scope of the present study is two-fold. First we examine how the
organization of native contacts impacts on fundamental properties of
proteins and, hence, to what extent the knowledge of the contact map
can be exploited to predict experimentally verifiable quantities. Our
second goal is to repeat the same analysis in the context of generic
disordered globular structures. From the comparison of the outcomes we
aim at finding clues about the criteria adopted by nature in the
selection of viable folds.

The vibrational spectrum at finite temperature allows to determine
typical relaxation times which, in turn, correlate with high accuracy
with experimental folding rates. This is rather surprising and
unexpected since the relaxation time refers to near equilibrium
situations whereas the folding process is far from equilibrium. 

Furthermore, the comparative study of analogous static and vibrational
properties of disordered globular structures, or otherwise random graphs,
and proteins shows that
the latter have very uncommon properties in terms of flexibility and
mechanical relaxation times that can be exploited to predict not only
folding rates but also the key sites of an enzyme and amino acid
vibrational amplitudes in partially folded states.  These distinctive
mechanical properties reflect the hierarchical assembly of protein
native states~\cite{Baldwin} and, in particular, can be traced back to
the presence and organization of secondary elements. Hence, they add
to the increasing evidence that the emergence of such motifs was
promoted by special criteria operated by nature to select viable folds
\cite{plato,Chothia2,chan90,Socci94,opthelix}.

The straightforward implementation of concepts presented here, makes
the model under consideration a useful tool to characterize the
folding properties of a protein, complementing alternative
theoretical strategies or more labor-intensive experimental
techniques.

\section{Theory}

The model under consideration incorporates two features that previous
theoretical investigations have proved successful for protein
modeling. In spirit, it belongs to the class of Go-models~\cite{Go},
since the energy scoring function introduces a bias in structure space
favoring the formation of native contacts. This ensures that the
native structure under examination has minimum effective
energy. Furthermore, we make use of the observation that, near
equilibrium, the dynamics resulting from complicated atomic
interactions can be well-reproduced by simple harmonic potentials
\cite{tir96,levitt85,go_gauss91} at least for time intervals shorter that
1 ns~\cite{Amadei93}. The starting point of our analysis is the
following effective harmonic Hamiltonian~\cite{gaussian} (for its
derivation, see Appendix~\ref{appendix}):

\begin{equation}
H = \sum_{i,j} {\bf r_i} L_{ij} {\bf r_j} + f(T)\ ,
\end{equation}

\noindent where ${\bf r_i}$ denotes the distance-vector of site $i$
from its position in the native structure, and $f$ is a term that only
depends on the equilibrium temperature. The entries of the symmetric
matrix, $L$, incorporate the elastic forces that act on each residue
to restore the native separation with its nearest neighbors in
sequence (effect of the peptide bonding) and with other amino acids in
interaction in the native state. The peptide bond is modeled by
springs whose associated Boltzmann weight is independent of
temperature. This reflects the little change of the peptide coupling
over the range of temperatures where unfolding/refolding transitions
are studied. In addition to this contribution, at variance with
previous approaches, the energy scoring-function includes
temperature-dependent interactions between the pairs of residues
contacting in the native state. The list of such interactions is
summarized in the contact map, $\Delta$, whose entries, $\Delta_{ij}$,
are 1 if residues $i$ and $j$ are in contact in the native state
(i.e. their C$_\alpha$ separation is below the cutoff $c=7.5$ \AA) and
0 otherwise. For simplicity we refer to these temperature-dependent
interactions as non-covalent, although they also act on consecutive
residues. The strength of such non-covalent springs is calculated
self-consistently according to the method described in Appendix
\ref{appendix}. This makes the model non-linear but still very
tractable. When the temperature of the model is zero (this
mathematically-convenient idealization corresponds to a physiological
temperature where the protein is stable in its native state), all
springs have the same strength. As temperature is switched on, this
strength is reduced by a deterministic amount that is larger for
springs where the largest stretching is observed (see Appendix
\ref{appendix}).

\section{Results and discussion}

\subsection{Vibrational properties of proteins and disordered structures}

\begin{figure}
\includegraphics[width=3.0in]{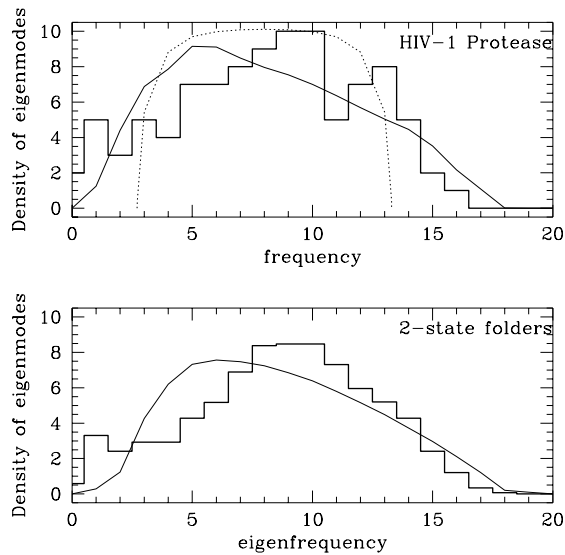}
\caption{(top) Density of eigenmodes for (a) HIV-1 Protease (heavy
line), (b) average over 3000 conservative reshufflements of HIV-1 PR
contact map (light line). The dotted line is the exact density of
eigenmodes for a regular graph with $k=8$ (see text). (bottom) Average
density of eigenmodes for the two-state folders listed in the caption
of figure~\ref{fig:fold_rate} (heavy line) and for 100 contact map
reshufflements for each of them (light line). Reported data refers to
near-native vibrations, corresponding to $T=0$ in our model. Previous
theoretical studies, carried out with full atomic detail, have shown
that the peak of the eigenmode distribution falls around a frequency
of 45 cm$^{-1}$ \protect\cite{levitt85,tirion93}.}
\label{fig:density}
\end{figure}

At zero temperature, when the non-covalent springs are equally strong
and dominate in number over the covalent ones, the model is equivalent
to the Gaussian Network Model (GNM) of Bahar {\em et
al.}~\cite{bah97,hal97,bah98}, which has been widely used to
characterize the mobile regions of the native state from the analysis
of the normal modes of the structure. These represent the natural
oscillations (independent of each other) of a protein and convey
information about which regions are more flexible, and how thermal
energy is dissipated in order to restore equilibrium (i.e. the native
structure). Although atoms in proteins are as tightly packed as in
crystalline solids, the vibrational properties of native structures
are very different~\cite{bav93}. A notable example is given by the
density of eigenvalues of the $L$ matrix, that is the histogram of
vibrational frequencies. This is shown in Fig.~\ref{fig:density},
where we concentrated on two examples: an individual enzyme, the
monomer of the HIV-1 protease, and the average histogram calculated
for several proteins known to fold via the simplest possible mechanism
(two-state)~\cite{Jackson}. The peculiarities of the protein spectrum
can be ascribed to several distinctive protein properties such as the
varying degree of site connectivity (burial profile) or the neat
hierarchical organization of the native state in terms of secondary
motifs.

To identify how the vibrational spectrum is affected by these general
features we shall not take the crystalline solids as a reference, but
consider several families of disordered contact maps.  For example, to
isolate the role played by the burial profile from other features we
first consider disordered matrices obtained by taking the contact map
of a real protein and randomly reshuffling its entries yet preserving
the symmetry of the matrix, the burial profile and the contacts
between consecutive sites. Successively we shall consider the case of
completely disordered maps but with equal connectivity (burial) for
each site.  It is important to stress that in these two cases of
disorder, the $\Delta$ matrices will not, in general, be physically
viable. In fact, it is not guaranteed that there exists a viable
three-dimensional backbone associated to an arbitrary symmetric
contact map. For this reason, we complete our analysis by considering
the case of contact matrices of three-dimensional compact
self-avoiding structures generated randomly with a computer with the
constraint to be as compact as naturally-occurring proteins. This case
will provide a useful term of comparison for identifying the spectral
properties resulting from the neatly organized three-dimensional shape
of naturally occurring protein conformations.

We begin by considering the first instance of disorder, i.e. random
reshufflings that preserve both the symmetry of $\Delta$ and also the
protein native burial profile, i.e. the native number of contacts to
which each site takes part~\cite{Miche99b}. The reshufflings consist
in picking randomly a pair of non-zero distinct entries, $\Delta_{ij}$ and
$\Delta_{kl}$ and checking if the entries $\Delta_{il}$ and
$\Delta_{kj}$ are zero. If this is so, the old pair of entries (and
their symmetric counterparts) are set to zero and the new ones to 1.
The preservation of the connectivity of each site allows to study the
impact of disorder on the vibrational frequency spectrum while keeping
burial profiles unaltered. As anticipated above, such randomized
contact maps are not necessarily the counterpart of any feasible
three-dimensional structure; in fact, a more appropriate view of such
matrices is within the framework of graphs~\cite{bollobas,Michefr},
(the nodes and links corresponding to beads and springs,
respectively).

A very important difference shown in Fig.~\ref{fig:density} is that
proteins have a larger number of low-frequency modes of vibration
compared to the spectrum of the reshuffled maps. This difference is
made more dramatic by considering the spectrum of loopless regular
graphs. The term {\em regular} here refers to the fact that all sites
have the same connectivity (burial), $k$, but the entries of the
associated symmetric matrix, $\Delta$, are otherwise
random~\cite{Merris94}.

This limit case of disorder is important because the average spectrum
of such ensemble of matrices has been calculated exactly by McKay
\cite{Mckay81}. The density of eigenfrequencies is given by:

\begin{equation}
f(\omega)=\left\{
\begin{array}{l l}
{k \sqrt{4 (k-1) - (\omega-k)^2} \over 2 \pi \ (k^2 - (\omega-k)^2)} & \ \mbox{for $|\omega-k| < 2 \sqrt{k-1}$}\\
0 & \mbox{otherwise,}
\end{array}
\right .
\end{equation}

\noindent and its distribution is shown in Fig.~\ref{fig:density} for
the case $k=8$, corresponding to the average site connectivity
(including peptide bonding) in real proteins (when the interaction
cutoff is around 7.5 \AA).

It appears that the range of vibrational frequencies is severely
limited compared to that of real proteins; in fact, the spectrum of
such graphs reproduces only the central part of the frequency
histogram of proteins. The outlying tails at both low and high
frequency are not captured at all. Previous studies have pointed out
how such tails reflect the existence of heterogeneity in site
connectivity (burial)~\cite{mona99,biro99}. This fact, and its
implications for the mechanical properties of natural biopolymers are
examined in detail in the next section.

\subsection{Localization properties}

\begin{figure}
\includegraphics[width=3.0in]{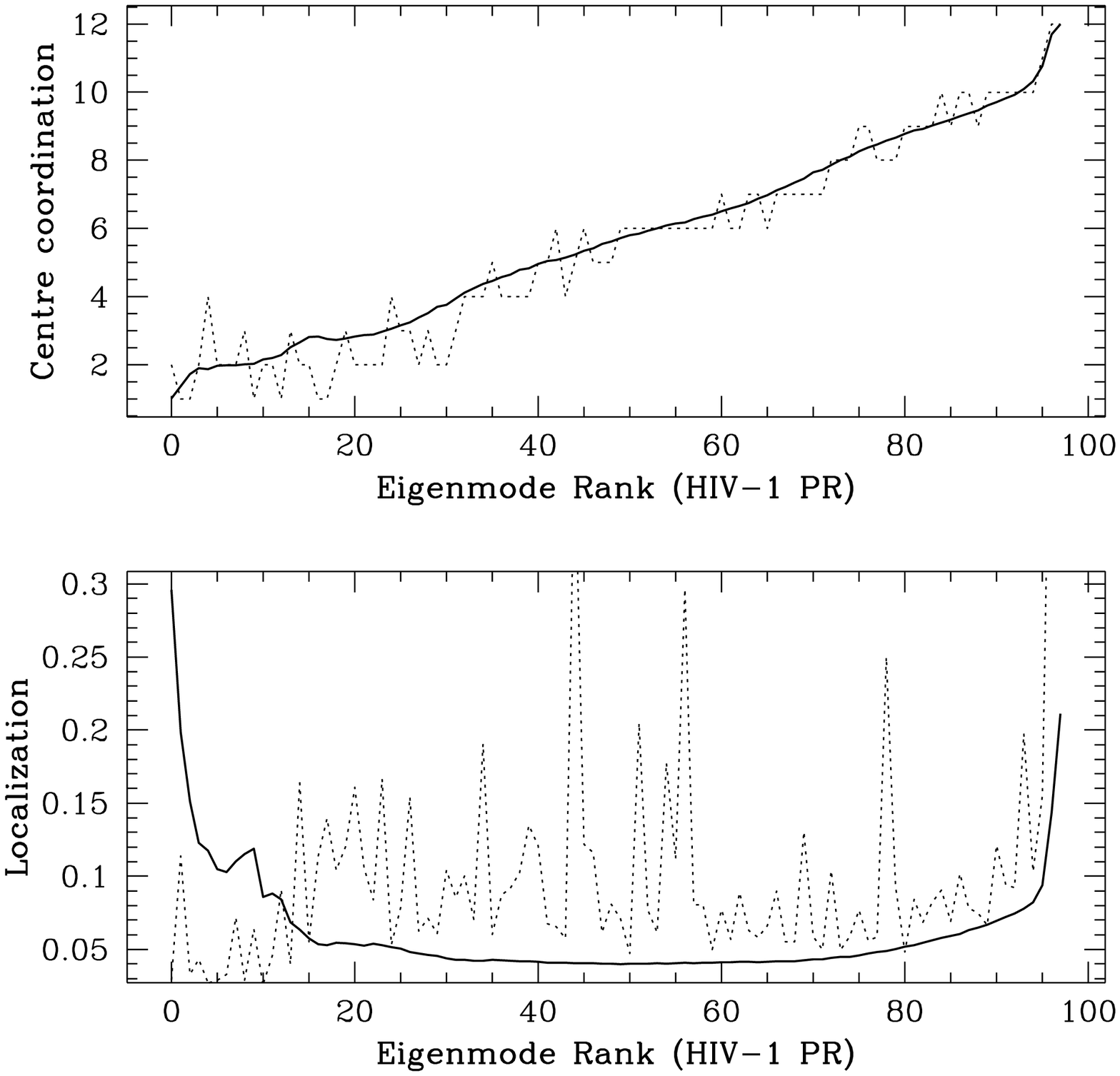}
\includegraphics[width=3.0in]{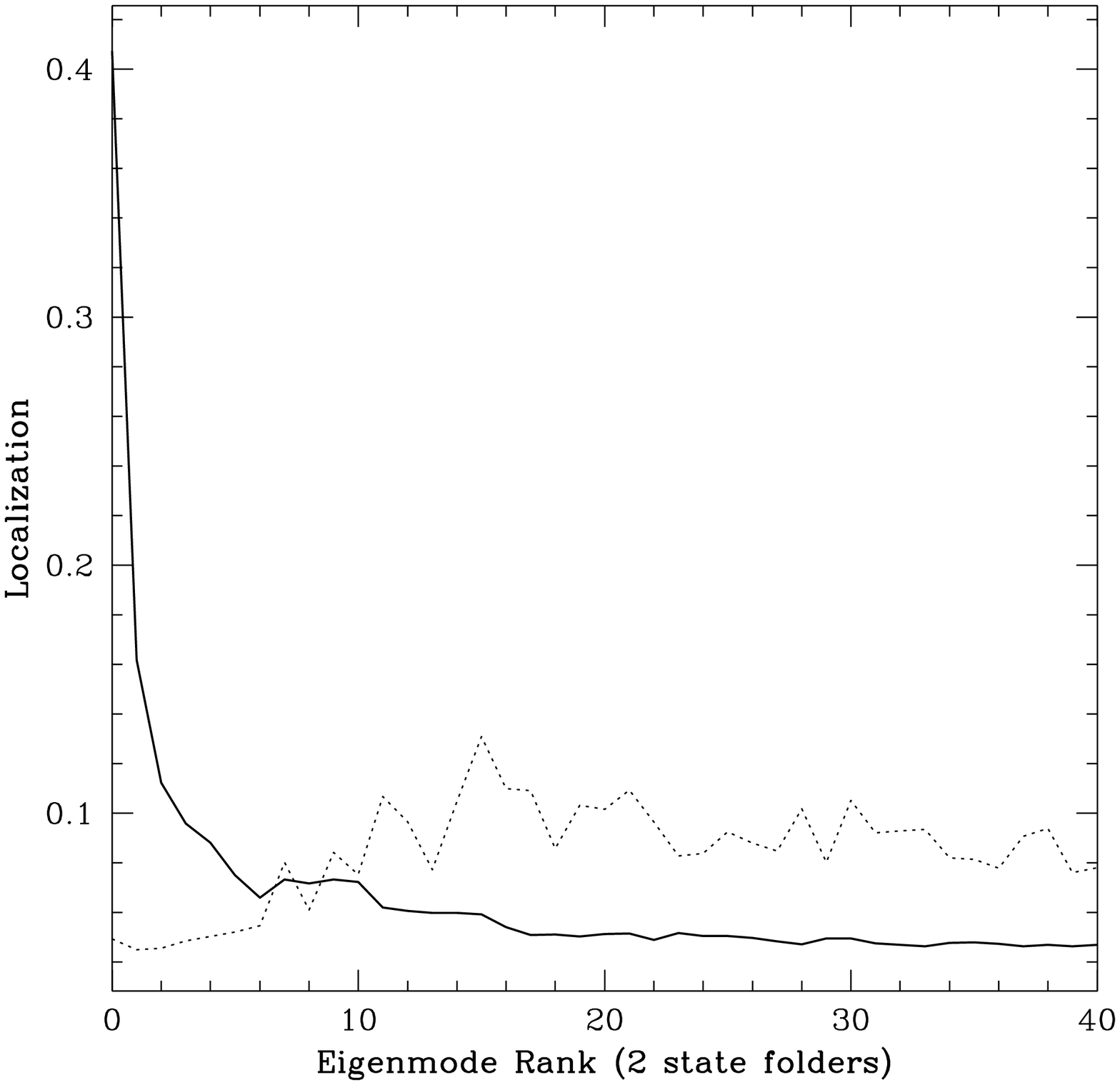}
\caption{(top left) Degree of burial (number of native contacts) of sites
that are at the center of the various eigenmodes. The eigenmodes are
ranked according to increasing vibrational frequency. The center of an
eigenmode is the site where one encounters the largest component (in
absolute value).(bottom left) Degree of localization of the eigenmodes. The
smooth solid lines pertain to averages over 3000 reshufflings of the
protease (monomeric) contact map (but preserving the site
connectivity). The dotted line relates to the native state.
(right) Degree of localization of the first 40 eigenmodes averaged
over the two-state-folders (dotted line) and over 100
reshufflements for each of them (solid line).}
\label{fig:burial}
\end{figure}

In a variety of contexts, from the physics of disordered systems to
graph theory, spectra have always attracted considerable attention
because of the special localization properties of the associated
modes. High frequency normal modes are usually concentrated on
sites/regions with high connectivity
\cite{And58,And78,mona99,biro99,levitt85,tirion93}. This is intuitive
since a larger number of connections leads to an enhanced local
stiffness and hence a higher frequency of vibration. Conversely, low
frequency modes are centered on sites/regions that, having fewer
connections than average, are more flexible~\cite{mona99,bah98}. This
overall picture applies remarkably well to the reshuffled contact
maps, as visible in Fig.~\ref{fig:burial}. The plots show that the
modes at low [high] frequency are very localized on sites with low
[high] connectivity (burial). The degree of localization of a
normalized eigenmode of vibration, $\{ v\}$ is defined as $\sum_j |
v_j|^4$ (see eg.~\cite{biro99}). This measure will take on the largest
value, 1, if all entries of $\{ v\}$ but one are zero (full
localization).  On the contrary, the localization measure will be
minimum when all the vector components are equal in modulus (full
delocalization).

It is now interesting to turn to the case of
proteins. Fig.~\ref{fig:burial} shows that high-frequency modes are,
as before, highly localized on heavily buried sites. Notice that the
highly localized states are not limited to the upper edge of the
spectrum. The picture, however, changes for low-frequency modes, which
appear to be significantly more delocalized than for the reshuffled
case. This denotes a degree of flexibility that is significantly
larger than randomly connected graphs (yet with the same burial
profile!). This is not a peculiar feature of the HIV-1 Protease
monomer, but is more general, as illustrated in the right panel of
Fig.~\ref{fig:burial} displaying averages taken over the two-state
folders listed in the caption of Figure~\ref{fig:fold_rate} and their
reshufflements. The averages in Fig.~\ref{fig:burial} are reported
only for the 40 slowest modes, which cover a similar range of
frequencies across the two-state folders. The inhomogeneity of the
lengths of such proteins prevents from taking straightforward averages
over the whole frequency range.

Several previous investigations of normal modes and spectra of
proteins~\cite{levitt85,tirion93,go_gauss91,bah97b,lus98,Hex98,rna98,jer99,hiv99,hal99,bah99,kes00,ani01}
had already noted the existence and utility of low-frequency modes in
proteins. For example the low-frequency spectrum affects not only the
mechanical stability of proteins but also the thermodynamic one
\cite{cieplak2001,hoang2000b,cieplak}; furthermore it has been argued
that the slowly moving regions are the natural candidates for
biological functions involving structural rearrangements. It is
important, however, to note that also the burial-preserving reshuffled
maps have low-frequency vibrations.

The novel message of Fig.~\ref{fig:burial} is that the slow modes of
random graphs are nearly fully localized (as high frequency modes),
while in proteins one encounters the slow movement of regions spanning
several residues \cite{levitt85}. This special property, which does
not depend on the burial profile, must result from the special
organization of native contacts which provides a remarkable degree of
flexibility. This, in turn, is associated with a rate of thermal
energy dissipation that is slower compared to disordered graphs. The
natural measure for the limiting rate at which mechanical excitations
decay is, in fact, given by the smallest frequency (eigenvalue) of
$L$. Equivalently, the slowest vibrational relaxation (decay) time in
the system, which we shall denote as $\lambda_0$, is given by the
inverse of the smallest vibrational frequency, $\lambda_0 = 1/\omega_0$.

\begin{figure}
(a)\includegraphics[width=0.45\textwidth]{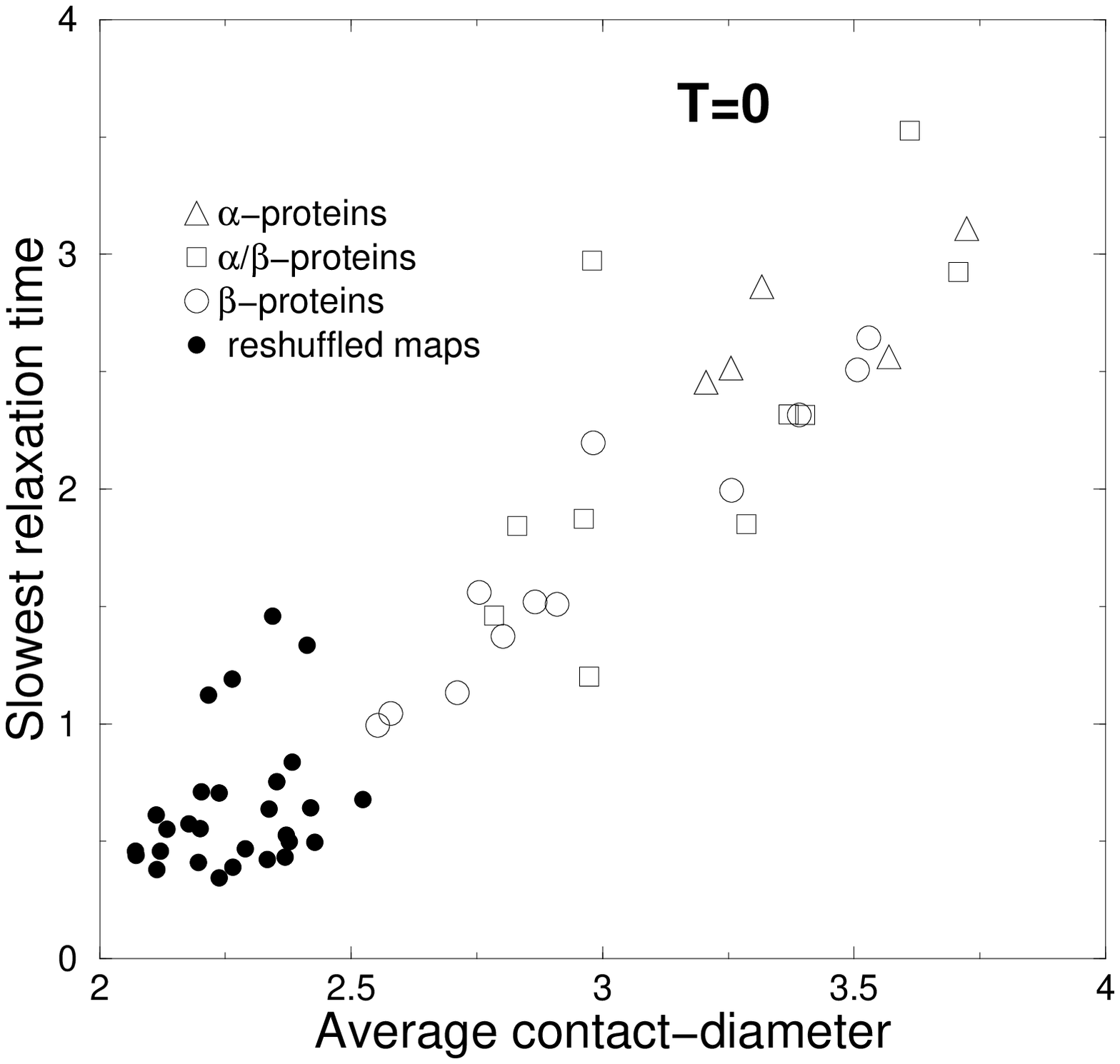}
\hfill
(b)\includegraphics[width=0.45\textwidth]{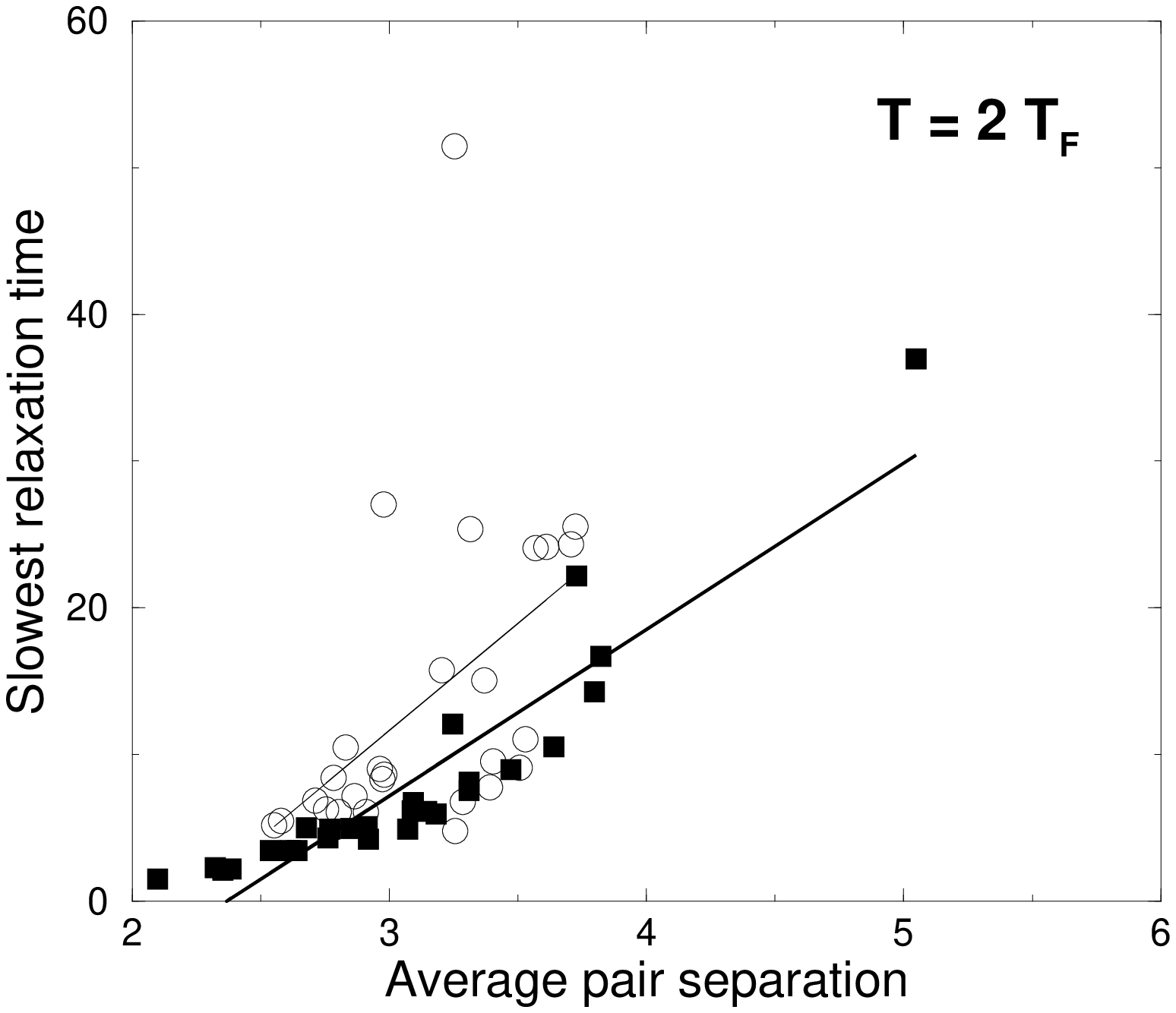}

\caption{(left) Scatter plot of the slowest native relaxation times
($T=0$) versus the average native contact diameter. The open symbols
denote data pertaining to the two-state folders listed in the caption
of Fig.~\ref{fig:fold_rate}. The open triangles, square and circles
denote proteins belonging to the $\alpha$, $\beta$ and $\alpha\beta$
families respectively.  The filled circles denote the slowest
relaxation time observed with a random (but burial-preserving)
reshufflement of the native contact map of each of the two-state
folders.  (right) Scatter plot of the slowest relaxation times at $T=
2\, T_F$ versus the average native contact diameter. The data pertain
to the two-state folders and an equal number of computer-generated
disordered compact structures with same length distribution (filled
squares).}
\label{fig:diameter}
\end{figure}

\noindent The immediate cause for this further difference is simply
described in terms of graph properties~\cite{bollobas}. There is a
well-defined relationship between the slowest relaxation time and the
average contact-diameter (meaning the average number of native
contacts, including those between consecutive sites, that have to be
traversed to go from an arbitrary site to another). This relation is
neatly visible in Fig.~\ref{fig:diameter}, which displays results for
the distinct two-state folders and an equal number of their
reshufflements.

It is important to see that, among all proteins, the ones with many
local contacts (all-$\alpha$ family) tend to be at one end of the
scatter plot, while proteins belonging to the all-$\beta$ family are
closer to the case of reshuffled contact maps. The relaxation of real
proteins appears to be also slower than randomly-collapsed
three-dimensional structures. Within our simplified approach this
difference is most visible at finite temperatures, as illustrated in
Fig.~\ref{fig:diameter}, where the relaxation rates are computed at
$T=2\,T_F$.

\begin{figure}
(a) \includegraphics[width=0.45\textwidth]{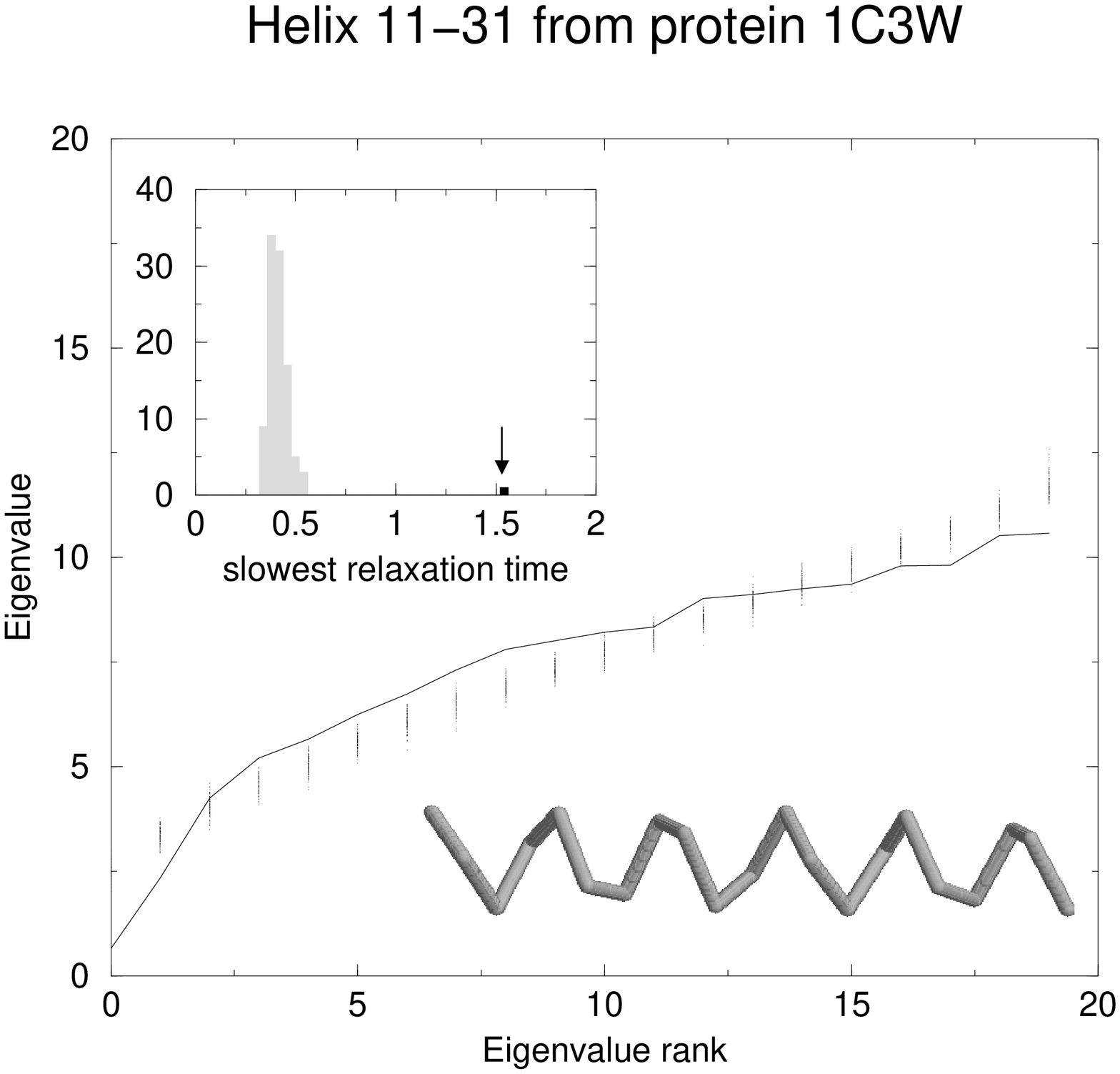}
\hfill
(b) \includegraphics[width=0.45\textwidth]{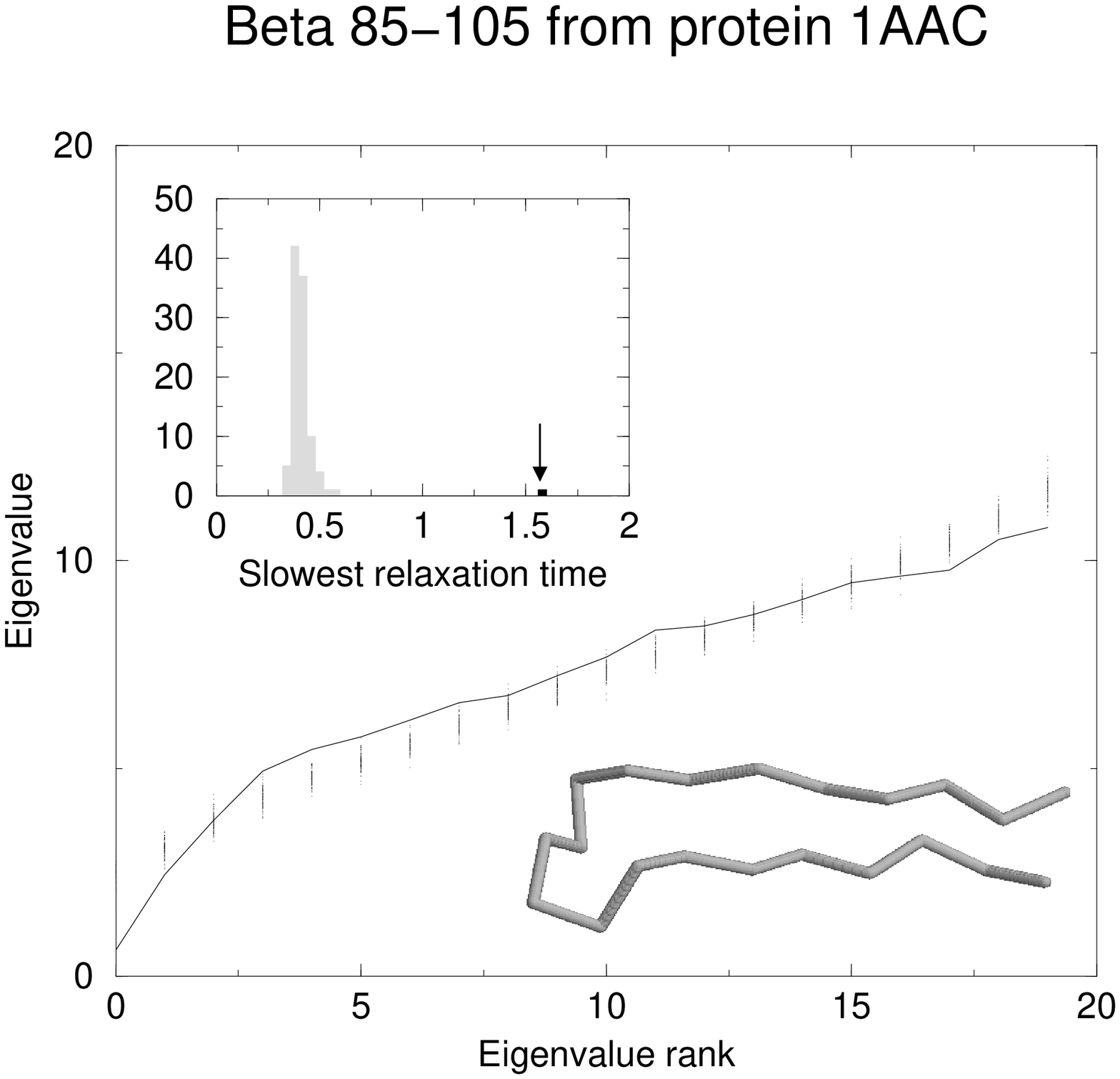}

\caption{Frequency spectrum at $T=0$ of (a) helix 11-31 from protein 1c3w
 and (b) the beta sheet 85-105 from protein 1aac
(continuous line). The dots denote the frequency spectrum calculated over 100
burial-preserving  random reshufflements of the original
contact map. The inset displays the histogram of the slowest
relaxation times, $\lambda_0$ for the reshuffled maps. The slowest
relaxation time for the original helix and beta sheet are denoted by
an arrow in the respective figures.}
\label{fig:helix-beta}
\end{figure}

The slowest relaxation encountered in proteins can be understood from
the fact that the establishment of a few random contacts, very
rapidly makes the structure rigid, and less susceptible to further
rearrangements. The hierarchical organization of contacts in proteins,
in terms of secondary motifs further arranged in the tertiary
structure, is responsible for the enhanced flexibility visible in
Fig.~\ref{fig:diameter}. In fact, the
relaxation times of {\em isolated} secondary motifs, both $\alpha$ and
$\beta$, of equal length is nearly the same and, again, much higher
than a corresponding random reshuffling. This is illustrated in
Fig.~\ref{fig:helix-beta}. From these facts it is tempting to
speculate that $\alpha$- and $\beta$-like motifs are among the
configurations that, for a given burial profile, have the largest
relaxation times. This is particularly intuitive in the case of
helices since their periodicity (modulus boundary effects) is
immediately associated to the presence of slowly-decaying excitations.

\subsection{Folding rates}

In this subsection we shall explore the intimate connection between
the relaxation time of slow eigenmodes of protein structures in
thermal equilibrium and the protein folding rate. It is tempting, and
physically appealing, to speculate that the relaxation rate of
partially folded states may be an indicator of the folding velocity.
This stems from the observation that low-frequency vibrations involve
extended protein segments and require less energy to be activated.
Hence, structural rearrangements from the unfolded state to the native
one should occur more easily when the relaxation times of partially
folded states are higher. This qualitative argument is also supported
by a different reasoning where the folding of a protein is viewed as a
diffusion in the complicated free-energy landscape until the
native-state (at the minimum of the accessible free energy) is
reached. Different native shapes will then be associated with
landscapes that, despite being biased towards the native conformation,
may have significant differences in terms of numbers of metastable
minima, height of the barriers between them etc. The corrugation of
the landscape will undoubtedly affect the folding rate.  The
relaxation time of a structure in thermal equilibrium reflects the
average ruggedness of the landscape. In a smoother landscape a protein
will be able to make larger conformational changes for the same amount
of dissipated energy. We can thus conclude that the native
state is reached with greater difficulty if, during the folding
process, the partially-formed structures have smaller equilibrium
relaxation times.

\begin{figure}[htbp]
\includegraphics[width=3.0in]{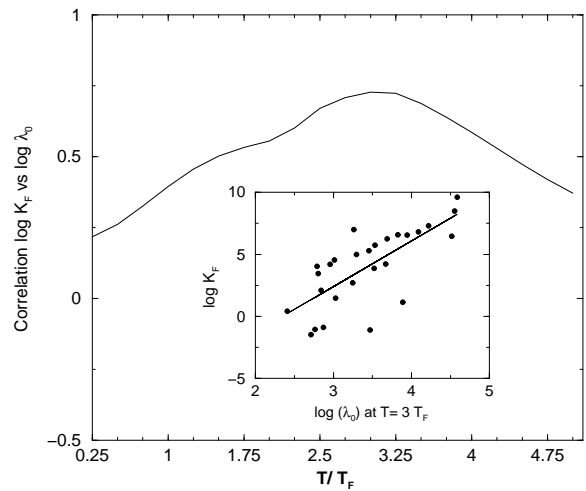}
\caption{Linear correlation coefficient between the logarithm of the
theoretical relaxation time, $\log \lambda_0$, and experimental
folding rates for a set of proteins known to fold via a two-state
mechanism. The inset shows the scatter plot of the actual theoretical
and experimental values at a temperature close to where the best
correlation, $r=0.73$ is observed. Notice that $K_F$ spans several
orders of magnitude. The folding rates were desumed from the
experimental studies collected in
\protect\cite{Jackson,Plaxco,goddard,karplusnsb,Finkelstein01,Michefr}. The
set contained the proteins listed below. $\alpha$ family: 1lmb, 2abd,
1imq, 1ycc, 1hrc; $\alpha/\beta$ family:
2gb1, 1div.n,2ptl, 1coa, 1hdn, 1div.c,1urn, 1aps, 1fkb, 2vik; $\beta$
family: 1shg,1srl,1shf.a, 1tud,1csp,3mef,2ait,1pks,1ten,1wit,1fnf
(9FN-III and 10FN-III).
When multiple experimental entries were available for the folding rate
of a give protein, we used considered the arithmetic average of $\log(K_F)$.}
\label{fig:fold_rate}
\end{figure}

In agreement with this picture, we have found a notable interdependence
(see Fig.~\ref{fig:fold_rate}) between the experimental folding rates,
$K_F$, and the slowest relaxation rate, $\omega_0$ (in
dimensionless units) which dominates the long-time relaxation kinetics
in equilibrium. At temperatures much larger than the folding
temperature, $T_F$ (identified as the temperature at which the peak of
the specific heat is observed), the relaxation time, $\lambda_0 = 1/ \omega_0$ conveys little
information about the folding rate, since virtually no contacts are
formed; this is true also at low $T$, where it measures only the rate
of thermal dissipation of the fully-folded native structure.

The highest correlation is observed at finite temperatures higher than
$T_F$ (see Fig.~\ref{fig:fold_rate}). Although we believe that the
precise location of the peak of the correlation may be sensitive to
the model details, this result is plausible since it is above the
folding transition that the significant search in conformation space
occurs. The correlation coefficient in the neighborhood of the optimal
temperature is $r=0.73$. It is possible to assess its statistical
significance by comparison with the null case of no correlation
between relaxation rates and folding times.  In the absence of any
pair correlation between variables with converging moments, one
expects that the correlation coefficient measured over a finite sample
is normally distributed.  This allows to calculate exactly the
probability to encounter a correlation coefficient higher than
$r=0.73$ if relaxation and folding times were statistically
independent.  This probability turns out to be equal to $2\cdot
10^{-4}$ which testifies that the observed dependence between the two
quantities is truly significant.

 These results add to previous studies where folding
rates were predicted with analogous level of significance on the basis
of contact locality~\cite{Plaxco} or different topologic indicators
(such as the clustering coefficient or cliquishness) based on a
graph-theoretical description of a protein's non-bonded
interactions~\cite{Michefr}.

 \subsection{B factors}

The choice of a simple Gaussian Network Model to investigate the
biological function of proteins has been motivated by a striking
qualitative agreement between the experimental $B$ factors (or
temperature factors) measured in X-ray diffraction experiments and the
mean square residue displacement calculated theoretically from the
expression~\cite{hal97,bah98,erm2002}:

\begin{equation}
B_i^{\rm theory} \propto \sum_k {1 \over \omega_k} | v_i^k|^2
\label{eqn:bfact}
\end{equation}

\noindent where $i$ is the residue index, $v_i^k$ is the $i$-th
component of the $k$-th mode whose eigenfrequency is $\omega_k$. The
sum in eqn. (\ref{eqn:bfact}) is over $k$ such that $\omega_k \neq
0$. Several alternative methods exist to predict (or refine)
experimental $B$ factors
\cite{levitt85,diamond90,tirion93,coarse2002,halle2002}. The appealing
feature of this Gaussian approach is that it is possible to give a
good account of the experimental $B$ factors by exploiting only the
connectivity information contained in the native structure.  A recent
work by Halle~\cite{halle2002} has remarked the heavy influence that
the native-state topology exerts on the measured $B$-factors; in
particular, it was shown that the mobility of an amino acid
anti-correlates with its degree of burial (estimated as the number of
non-covalent bonds to which it takes part). Besides the fundamental
effect of the burial profile, the GNM allows to model the influece on
$B$-factors of other important topogical features, such as the
locality of contacts and presence of highly-interconnected clusters of
interactions. In addition, the decomposition of dynamical motion into
independent modes allows to trace which modes of vibrations are most
responsible for the mobility of a given site or protein region, thus
providing useful hints about the protein biological function.

In this section we perform an analysis of the correlation between the
theoretical and experimental B factors, aimed at investigating the
role played by the single vibrational eigenmodes in the overall observed
temperature factors. 

We stress that in our model, both eigenfrequencies and eigenmodes are 
temperature-dependent. We begin our analysis by considering the $T \to 0$ 
limit of our model, that corresponds to the GNM studied in 
refs.~\cite{bah97,hal97,jer99}.

\begin{figure}
\includegraphics[width=3.0in]{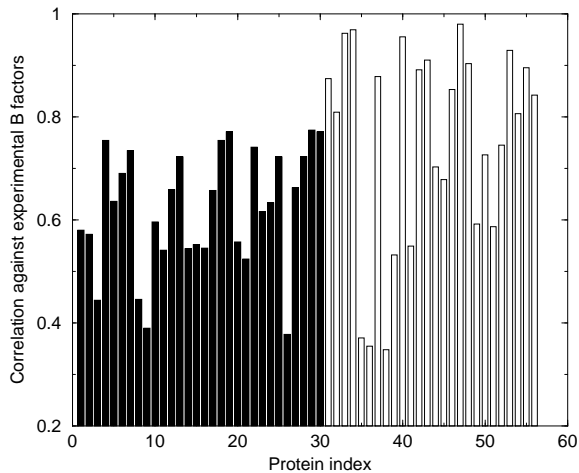}
\caption{\label{fig:corrB} Correlation between the theoretical $B$
factors and the experimental ones reported in 30 X-ray resolved
proteins (filled bars) and NMR resolved ones (open bars). The proteins
in the two classes were ordered according to increasing lengths (the
overall length range is 39-258). The PDB code of the proteins adopted
here are reported below. {\bf X-ray} (1 to 30):1shg,  1fxd, 1igd,
1ail, 1hoe, 1vcc, 1cei, 1opd, 1fna, 1pdr, 1beo, 1poa, 1mai, 1bfg,
1laa, 1kuh, 1cof, 1lcl, 1pkp, 1vsd, 1npk, 1vhh, 1gpr, 1sfe, 1amm,
1ido, 1kid, 1cex, 1chd, 1thv; {\bf NMR} (31 to 56): 1aqr, 1aa3, 1ikl,
1bo0, 1awj, 1a8c, 1a1w, 1bhu, 1ocd, 1ght, 1trs, 1jli, 1fwq, 1tam,
1ahk, 1tbd, 1gdf, 1joo, 1fls, 1buy, 1il6, 1aa9, 1ak6, 1a23, 1lxl,
1eza.}
\end{figure}

We analyzed 30 protein structures, all obtained in X--ray diffraction
experiments, of different size (up to about 200 amino acids). The
proteins, listed in the caption of Fig.~\ref{fig:corrB}, are chosen
among the monomeric representatives of distinct structural classes
obtained from high-resolution data~\cite{oligons}. Thermal
fluctuations for each amino acid were compared with the corresponding
temperature factor reported for the CA atoms in the PDB file
\cite{pdb}. Although a remarkable agreement with experimental data can
be obtained for several proteins (correlation coefficients larger than
0.75 for proteins with nearly 200 residues), in some rare instances
the correlation was around 0.4 (see Fig.~\ref{fig:corrB}). It must be
borne in mind, however, that such correlations are never trivial,
since they are measured over the entire protein length. Assuming, as
before, that in case of no correlation the distribution of $r$ is
normal, we can calculate the probability to observe higher
correlations than the measured ones. Even in the worst case, this
probability is never larger than $t=3\ 10^{-4}$ which is an indicator
of excellent correlation. This estimate of statistical significance is
so small that even if more precise calculations (e.g. accounting for
the skewness in the B-factors distributions) modify the estimate of
$t$ by two orders of magnitude, the correlation significance would be
high.

Although statistically relevant, the agreement with temperature factors
does not have the same quality for all crystal structures. It is important 
to remark that the crystallographic
$B$-factors are not obtained by direct measurement, but from a
suitable parametric fit of the measured diffraction pattern. In
unfavorable conditions (affected also by crystallographic
resolution), the fit may be significantly underdetermined and reliable
data can be obtained only imposing additional constraints on the $B$
factors (such as their smooth variation along the main- or
side-chains). Inspection of the PDB files for which the worst
correlations are seen have revealed that, indeed, those proteins
lacked the $B$-factor smoothness. Another possible source of
discrepancy between theoretical and experimental data is the fact that
we neglected the proximity effect of the surrounding proteins in the
crystal structure. The presence of neighboring proteins might reduce
the mobility of some regions (and notably those experiencing a pronounced 
freedom of movement), as already noted in previous studies
\cite{levitt85,tirion93,coarse2002,halle2002}.

\begin{figure}
\vskip 0.7cm
(a) \includegraphics[width=0.4\textwidth]{fig7a.eps}
(b) \includegraphics[width=0.4\textwidth]{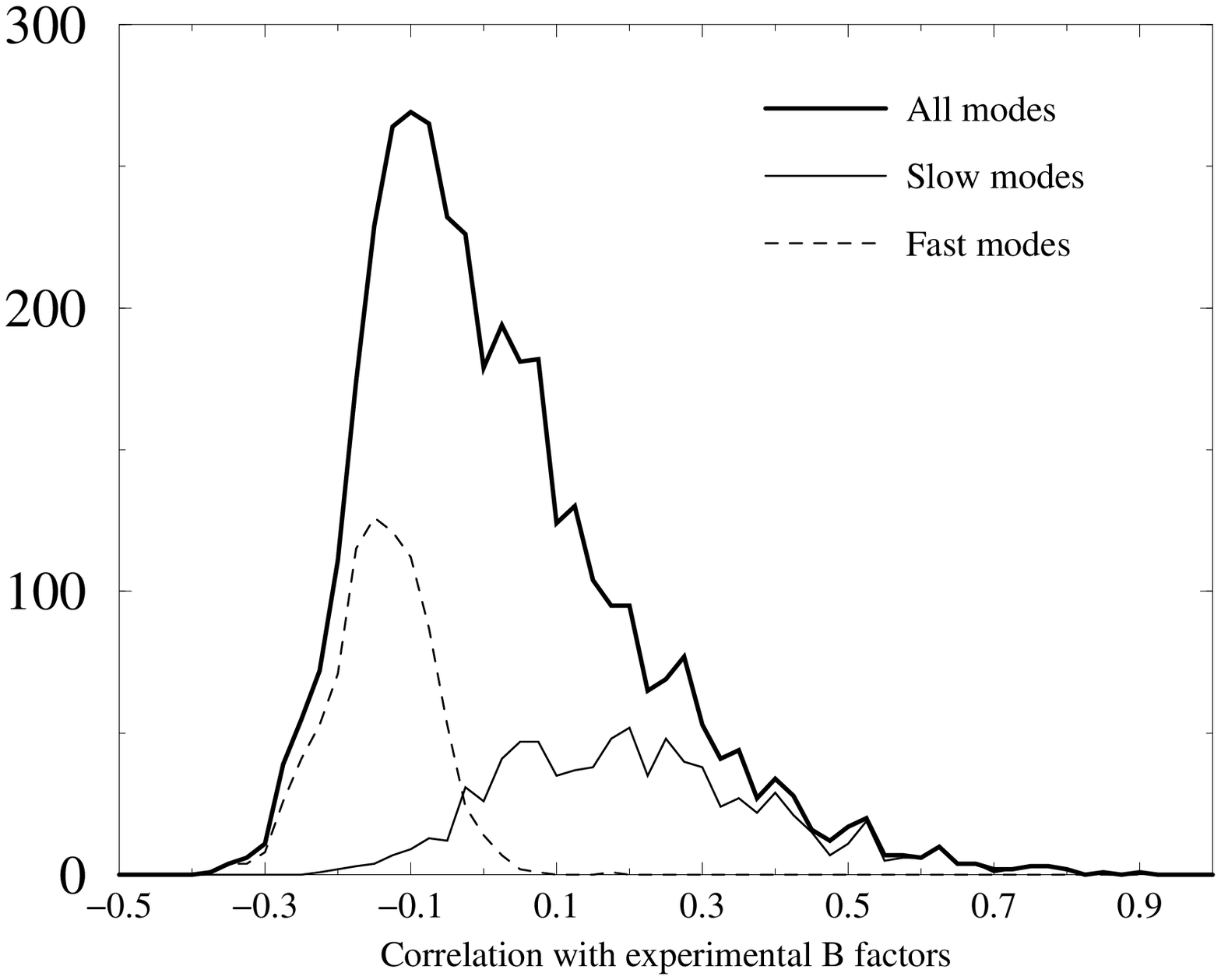}

\caption{\label{fig:single}(a) Correlation coefficients between single
eigenmodes amplitudes (only one term at a time is considered in
eqn. (\ref{eqn:bfact})) and experimental $B$ factors as a function of
their corresponding eigenfrequencies. The points refer to all
eigenmodes of the X-ray structures listed in the caption of
Fig.~\ref{fig:corrB}.(b) Histograms from data in (a). Positive
correlations come from slow modes, negative from fast modes. }
\end{figure}

Equation (\ref{eqn:bfact}) clearly indicates that most of the
contribution to the theoretical $B$ factor comes from low-frequency
modes, since the fast ones are suppressed by their frequency
reciprocal weight. It is sensible to ask how much each {\em
individual} mode correlates with the experimental $B$ factors. If a
clear trend of such correlation against the mode frequency is found,
one could devise a better (knowledge-based) weighting scheme,
alternative to the one of eqn.~\ref{eqn:bfact} to improve the
theoretical estimates of mean square displacement of residues. To
answer this question we have obtained for each mode, $k$, the linear
correlations between the experimental $B$ factors along the chain,
$\{B^{exp}_i \}$ and the mode amplitudes, $\{ |v_i^k|^2 \}$. Such
correlations, computed separately for each mode of the proteins listed
in Fig.~\ref{fig:corrB} are reported in Fig.~\ref{fig:single}a.

The majority of positive correlations are found in the region of
slowest eigenmodes (eigenfrequencies up to 6), while there is a
negligible correlation for middle frequencies, and a
weak anti-correlation for fast eigenmodes (frequencies larger
than 12). This is evident in figure~\ref{fig:single}b
where one can compare the histograms of the correlation coefficients
cumulated (with equal weights) over fast and slow eigenmodes. The
anti-correlation of the experimental $B$-factors with the
square-amplitudes profile of fast eigenmodes is in agreement with the
graph theoretical prediction of a localization of the fastest
eigenmodes in nodes of higher connectivity. These nodes are less
mobile in the protein structure and hence associated with smaller $B$
values. Furthermore, since all eigenmodes are orthogonal, the sites
whose vibrational amplitude is large in the slowest eigenmodes, will
not be very mobile in the fastest ones, and viceversa. 

Following this analysis we can conclude that low eigenmodes 
cooperate to generate much of the fluctuation pattern of the protein
structure, while intermediate modes have little influence and fast
modes could be safely excluded from the sum, since they contribute to
raise the mobility of highly connected nodes.

\begin{figure}
\includegraphics[width=3.0in]{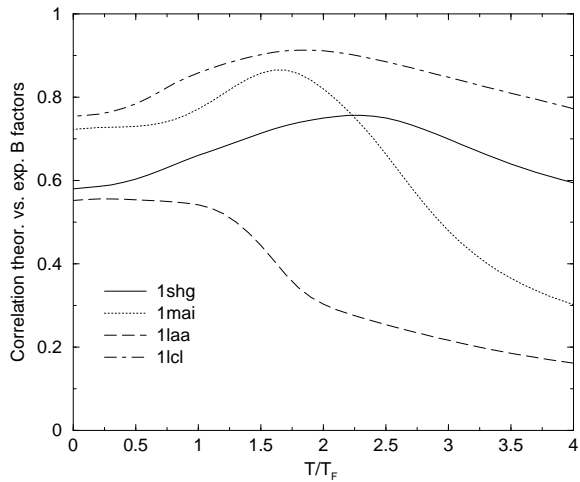}
\caption{\label{fig:Bft} Correlation between theoretical and
experimental B factors as a function of temperature.}
\end{figure}

Finally, we observe that the theoretical $B$'s carry a non trivial
dependence on temperature through the eigenmodes (due to the
temperature dependence of matrix $L$). Fig.~\ref{fig:Bft} summarizes
the behaviour of the correlation with experimental data against
temperature for four representative proteins. It can be seen that,
typically, by working at a temperature around $2\, T_F$ the
correlation with experimental data can be increased significantly. As
visible, an exception to this trend is given by the few proteins which
already have a poor correlation at $T=0$.

\noindent We have considered also the case of NMR proteins, for which
accurate relaxation measurements can directly probe the mobility of
local portions of the protein thus providing an {\em effective} $B$
factor.  Strikingly, the correlation between the theoretical and the
experimental $B$ factors observed in this situation was systematically
higher than for the X-ray resolved counterparts, as can be ascertained
from Fig.~\ref{fig:corrB}. It would be tempting to conclude that this
improved agreement is due to the more direct way with which NMR
experiments can address local vibrational amplitudes. However, due to
the insufficient annotation of PDB files of some NMR-resolved
structures, it is not always possible to ascertain whether the
reported data follow from precise relaxation measurements or from an
{\em a posteriori} analysis of the various model structures compatible
with measured distance-constraints.

\subsection{B factors and key sites}

The analysis of the mobility, in thermal equilibrium, of different
protein residues carries significant information about sites that are
important in the folding process. We illustrate this important
application for the case of the HIV-1 Protease. This enzyme represents
an ideal benchmark due to the vast number of clinical studies thanks
to which comprehensive tables of key mutating sites have been compiled
\cite{Condra,BIOCH88,gulnik,hiv1,apr,tisdale}. Molecular dynamics
studies have attempted to clarify the special role played by these
amino acids either through all-atom simulations of the protease
\cite{piana2002} or through topological folding models \cite{hiv}. It
appears that a number of key mutating sites form contacts that act as
rate limiting steps for the folding process~\cite{hiv}.

One attempt to describe/predict this crucial set of amino acids from
the analysis of normal modes was carried out in
Refs.~\cite{bah97,hal97,jer99} within GNM (to which our model reduces
when $T=0$), who pointed out that high frequency modes are localized
close to key mutating sites~\cite{hiv99}. Our results, especially
those of Fig.~\ref{fig:burial}, support the view that the sites where
the high frequency modes concentrate are paramount for native
stability; in fact they are among the least exposed ones. However,
this property is only related to the native structure, and does not
imply that the same sites play the leading role in overcoming the
folding rate-limiting steps.

\begin{figure}
\includegraphics[width=3.0in]{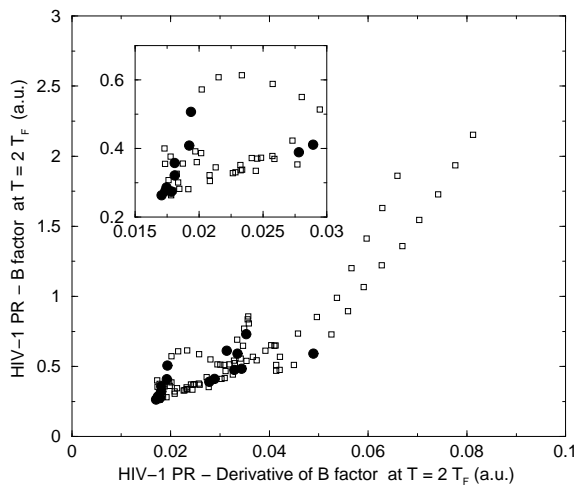}
\caption{Scatter plot of B-factors and their temperature derivative
for the sites of HIV-1 PR monomer. The key sites known to cause
resistance to protease-inhibiting drugs are highlighted.}
\label{fig:key1}
\end{figure}

 A more direct strategy would be to determine the handful of contacts
that form cooperatively at, or in the neighborhood of, the folding
transition temperature. The formation of such contacts, which has an
all-or-none character, is expected to constrain significantly the
mobility of the residues involved in their establishment. This scheme,
which has been confirmed {\em a posteriori} in simplified folding
models of the HIV-1 PR and prion~\cite{hiv,goprion}, can be naturally
adopted in the present case. In fact, a natural and convenient
measure of the degree of spatial constraint of a given site is
provided by the associated B-factor. Thus, near the folding transition
temperature, one expects that the key sites have nearly native like
values for the $B$ factors, while other residues will have much larger
fluctuations than in the native state. Therefore, one may identify the
key sites as those with small values of $B$ or small derivatives of
$B$ with respect to temperature. It turns out that both these criteria
work well, as illustrated in Fig.~\ref{fig:key1}. Among the top 13
sites ranked according to the smallness of $B$ factors there are 5 key
residues: 32,33,77,84,30. In the top 13 sites ranked according to the
derivative of $B$ there is an additional key site, namely residue
82. The probability to observe at least these many matches had we
chosen the sites randomly, would have been 3 \% and 0.5 \% respectively
in the two cases. It must be stressed that the present strategy to
identify key sites exploits the information on native topology in a
non-trivial manner avoiding, for example, the pitfall of selecting the
most buried sites as the key ones. In fact, to collect the same
number of correct matches found before, one needs to retain about
twice as many burial-ranked sites than for the B-factor-ranked
case. We have verified that this conclusion holds for all interaction
cutoffs in the range 6.5-8 \AA. This very good statistical validation
shows that the physically appealing identification of the key sites
through normal modes analysis can be fruitfully applied in contexts of
high practical importance.

\section{Conclusions}

We have focused on the characterization of both the folding process
and equilibrium physical properties of natural proteins through the
normal modes analysis of a suitable topological model. Previous
investigations of normal vibrations around the native state have
proved a valuable tool to describe the large scale motion of proteins
and obtain clues about their biological functionality.

Although we also focus on specific protein instances, such as the
HIV-1 protease, our scope is more directed at finding some general
characteristics of proteins' dynamics in thermal equilibrium (at
various temperatures) that distinguish these biopolymers from a
randomly compact polymer. This comparison has a two-fold objective:
one of theoretical interest, the other more practical.

>From the point of theoretical biophysics, the identification of
features that are unique (and common) to proteins provides clues about
the special evolutionary criteria that have promoted the selection of
the wide repertoire of naturally-occurring proteins yet using only a
limited number of fold families~\cite{plato,Chothia2,secstr}. In the
detailed account provided here we have shown that proteins are very
flexible and have long relaxation times, in thermal equilibrium,
compared to globular polymers. In turn, the relaxation time of
partially folded states is shown to have a significant correlation
with experimental folding times ($r$ can be as high as 0.73). We give
a quantitative account of the intimate dependence of the relaxation
rates with native state topology; thereby providing a novel additional
support to the influence of native-state topology over folding rate.
It is also found that high-frequency modes involve as many sites in
natural proteins as in disordered structures. Despite this, the
structural regions involved in slow motions are much more extended in
proteins. Without this property (common to the few tens of proteins
considered here) it would probably be impossible for a protein to
carry out the articulated mechanical tasks necessary for biological
functionality~\cite{bah99,hal99,hiv99,kes00}. The direct
investigation of the vibrational properties of individual secondary
motifs has shown that their presence and organization are distinctive
features of protein structures. This observation adds to previous
evidence according to which the ubiquity of secondary motifs in real
proteins is due to the very special properties that they confer both
to the native state and to the folding process.

>From a more practical point of view, the knowledge of how these
"special features" have arisen through the selection of certain viable
native shapes, can be used as a predictive tool. In particular, the
straightforward deterministic analysis of the topologic folding model
adopted here allows to predict folding rates with high statistical
significance. A further important application is in the "a priori"
determination of the B factors that are essential in the refinement
and validation of protein structures resolved by X-ray and NMR.
A by-product of the analysis of B factors of partially
folded states is the identification of the sites (or contacts) that
lock into their relative native position at early stages of the
folding process. Taking the case of HIV-1 PR as a reference, we have
shown that such sites are among those that are known to be involved in
rate-limiting steps of the folding process. Therefore, within this
single simple framework it is possible to model various aspects of the
folding process and other equilibrium properties obtaining a thorough
protein characterization that is consistent with available
experimental results.

This work was supported by INFM and MURST cofin 2001. GL's research
has been supported by a Marie Curie Fellowship of the European
Community Programme ``Improving Human Research Potential and the
Socio--Economic Knowledge Base'' under contract number
HPMF--CT--2001--01432. The authors are solely responsible for
information communicated and the European Commission is not
responsible for any view or results expressed.  We are indebted to
Jayanth Banavar, Ken Dill, Burak Erman, Doriano Lamba and Henriette
Molinari for stimulating discussions and helpful suggestions.

\appendix
\section{Appendix. Methods: Analytical characterization of the model}
\label{appendix}

The simplified energy functional used to characterize the equilibrium
or folding properties of a target structure, $\Gamma$ is:
\begin{equation}
{\cal H} = \frac{K \cdot T}{2} \sum_{i=1}^{N-1}
({\bf r}_i- {\bf r}_{i+1})^2 + \frac{1}{2} \sum_{i \neq j}
\Delta^\Gamma_{ij} [({\bf r}_i- {\bf r}_{j})^2 - R^2]\, p_{ij}
\label{eq:gauss}
\end{equation}

\noindent where ${\bf r}_i$ denotes the distance-vector of the
$i$th $C_\alpha$ from its position in the native structure,
$\Gamma$ (assumed to be $N$ residues long). $K$ is the strength of the
peptide bonds, and $T$ is the absolute temperature (incorporating the
Boltzmann constant). $\Delta^\Gamma$ is the contact matrix, whose
element $\Delta_{ij}$ is 1 if residues $i$ and $j$ are in contact in
$\Gamma$ (i.e. their C$_\alpha$ separation is below a cutoff $c$. (in
this study $c= 7.5$ \AA) and 0 otherwise. The matrix
$\Delta^\Gamma_{ij}$ along with the native $C_\alpha$ coordinates
encodes the topology of the protein. In standard off-lattice
approaches, the interaction $V(d)$ between non-bonded amino acids at a
distance $d$, is taken to be a square well potential, or some type of
Lennard-Jones interaction. Our choice in Eq. (\ref{eq:gauss}) is a
sort of ``harmonic well'' which, while being physically sound and
viable, is suitable for a self-consistent treatment, as explained
below.

The temperature-dependent term $p_{ij}$ is the probability that the
separation of amino acids $i$ and $j$ fall within a distance $R$ of
the native separation. Therefore $R$ plays the role of an effective
outer rim of the quadratic potential well and can be set to a few
Angstroms ($R = 3$ \AA\ in the present study) to reflect the fact
that, when the separation of two residues exceeds substantially the
native one their interaction is negligible. The strength of the
peptide bond, $K$, is set to $1/15$ \cite{gaussian}. Such choice of
values guarantees that close to the folding transition temperature,
nearly half of the native contacts are formed, consistently with
several unrelated studies \cite{karp,Jackson}.

The simple quadratic form of Hamiltonian in Eq.~(\ref{eq:gauss}) allows to
determine exactly the $p_{ij}$'s from the self-consistent relation:

\begin{equation} p_{ij} = \langle \Theta(|{\bf r}_i - {\bf r}_j|^2 - R^2) \rangle.
\label{eq:self}
\end{equation}

\noindent where $\Theta(x)$ is the unitary step function and the
brackets denote the thermal averaging under the action of Hamiltonian
\ref{eq:gauss}, that depends on $p_{ij}$ as well. Operatively, one
starts from a trial choice of the $p_{ij}$, which is used to determine
a new set of parameters through eqn. (\ref{eq:self}). Convergence is
obtained in a few tens of iterations at any temperature. Now in such
self consistent approach the problem is fully solved and all
equilibrium or dynamical quantities can be calculated exactly by
evaluating analytical expressions. At $T=0$ the model correctly
assigns the lowest energy to the native structure (all $p_{ij}$'s
equal to 1). Upon increasing the temperature, each interacting pair
will have increasing mutual separation (in modulus) and
correspondingly, the $p_{ij}$'s will decrease to reflect the milder
binding. In the limit $T \to \infty$ all $p_{ij}$'s tend to 0.  It is
worth pointing out that our energy functional treats non-native
contacts in a neutral way: their formation is not favoured but is not
discouraged either. As a consequence, at finite temperatures,
non-native contacts have a non-zero probability of formation, which
can still be computed through~\ref{eq:self}.

Simple algebraic operations allow to recast the self-consistent energy
function in the following form:

\begin{equation}
{\cal H} = \sum_{ij} {\bf r_i} L_{ij} {\bf r_j} -
{1 \over 2} \sum_{ij} \Delta_{ij}^\Gamma\, R^2\, p_{ij}
\end{equation}

\noindent where

\begin{equation}
L_{i,j} = \left \{ 
\begin{array}{r l}
K\cdot T\, (2 - \delta_{i,1} - \delta_{i,N})/2 + \sum_l \Delta^\Gamma_{i,l} p_{i,l} \
\ & {\rm for }\ i=j \\
- p_{i,j} \Delta_{i,j} + K \cdot T\, \left( - \delta_{i,j+1} - \delta_{i,j-1}\right)/2 \ \ & {\rm for }\ i \not= j \ . 
\end{array}
\right.
\end{equation}

While the present form of the model does not accurately describe the
effects of self-avoidance this does not lead to a qualitatively wrong
behaviour in the highly-denatured ensemble (large $T$). The treatment
of steric effects becomes progressively more accurate as temperature
is lowered. In fact, the model guarantees that the native state is the
true ground state and therefore protein conformations found at low
temperature inherit the native self-avoidance. The connectedness
of the chain, as well as its entropy, are captured in a simple but
non-trivial manner. The most significant advantage of the model is
that it can be used to explore the equilibrium thermodynamics without
being hampered by inaccurate or sluggish dynamics.

\end{document}